\documentclass[twocolumn]{aastex62}
\usepackage{amsmath}

\graphicspath{{./}{figures/}}

\newcommand{\hnot}{63.01 $\pm$ 1.79~km s$^{-1}$ Mpc$^{-1}$}

\shorttitle{The Moon and $H_{0}$}
\shortauthors{Anand, Claytor, $\&$ Dungee}

\begin{document}

\title{\textbf{Worry No More, The Hubble Tension is Relieved: \\ A Truly Direct Measurement of the Hubble Constant from Mooniversal Expansion}}

\author{Gagandeep S. Anand}
\affiliation{Space Telescope Science Institute, 3700 San Martin Drive, Baltimore, MD 21218, USA}
\thanks{All authors contributed equally to this manuscript.}

\author{Zachary R. Claytor}
\affiliation{Institute for Astronomy, University of Hawai‘i, 2680 Woodlawn Drive, Honolulu, HI 96822, USA}
\thanks{All authors contributed equally to this manuscript.}

\author{Ryan Dungee}
\affiliation{Institute for Astronomy, University of Hawai‘i, 640 North A‘ohōkū, Hilo, HI 96720, USA}
\thanks{All authors contributed equally to this manuscript.}

\date{1 April 2022}

\begin{abstract}
Using sedimentary and eclipse-based measurements of the lunar recession velocity, we derive a new local-Universe measurement of the Hubble constant ($H_0$) from the recession rate of Earth's Moon. Taking into account the effects of tides, we find a value of $H_{0} = $ \hnot, which is in approximate agreement with the \textit{Planck} space mission's measurement using the cosmic microwave background (CMB) and base $\Lambda$CDM. Our new measurement represents the first ever model-independent, single-step measurement of the Universe's current expansion rate. This is also the first major local Universe measurement of $H_0$ which is \textit{below} the measurement from \textit{Planck}. Importantly, it is robust to the systematic errors that may be present in other $H_0$ measurements using other cosmological probes such as type Ia supernovae, baryon acoustic oscillations, or lensed quasars. Our work provides key evidence towards the notion that the existing Hubble tension may indeed be a result of systematic uncertainties in the local distance ladder.
\end{abstract}

\section{Motivation} \label{sec:intro}

Cosmology today is at an important juncture. There is a large amount of discussion on the value of the Hubble Constant ($H_{0}$), which denotes the present expansion rate of the Universe. Measurements of the Cosmic Microwave Background (CMB) from the Planck Satellite provide an extremely precise value of 67.4 $\pm$ 0.5 km/s/Mpc \citep{Planck2020}. On the other hand, the latest results from the most prominent ``local" measure of the Hubble constant is from the SH0ES team, who use observations of Cepheids in 42 nearby galaxies to derive of value of $H_{0}$ = 73.0 $\pm$ 1.0 km/s/Mpc \citep{Riess2022}. These two measurements are in 5$\sigma$ disagreement with each other, a fact which has sparked an incredible amount of interest within the community \citep{Schoenberg2021} to come up with a possible solution to this Hubble ``Tension". 

Other local Universe measurements also support values of $H_{0}$ which are higher than that obtained from Planck, with many different techniques being explored to provide a cross-check on the Cepheid results (although some of these measurements are somewhat consistent with Planck, such as those from the Tip of the Red Giant Branch). These techniques include Mira variables \citep{Huang2020}, Surface Brightness Fluctuations \citep{Blakeslee2021}, the Tip of the Red Giant Branch (TRGB, \citealt{Freedman2019, Anand2021}), and the Tully-Fisher relation \citep{Kourkchi2020}, among many others (see \citealt{2021APh...13102605D} for a more complete list).

It has been previously pointed out that the Moon's recessional velocity is on the same order as the implied Hubble flow expansion in the Earth-Moon system \citep{Maeder2021}, although the origin of this coincidence is unclear. In this work, we combine this suggestion with improved measurements on the Moon's recessional velocity and make the assertion that the origin of the Moon's recessional velocity (after discounting the known effects of tides) is from the Hubble expansion. In doing so, we provide a precise local measurement of the Hubble constant from the first \textit{truly} local measurement of $H_{0}$.

\begin{figure*}
    \centering
    \includegraphics[width=0.95\textwidth]{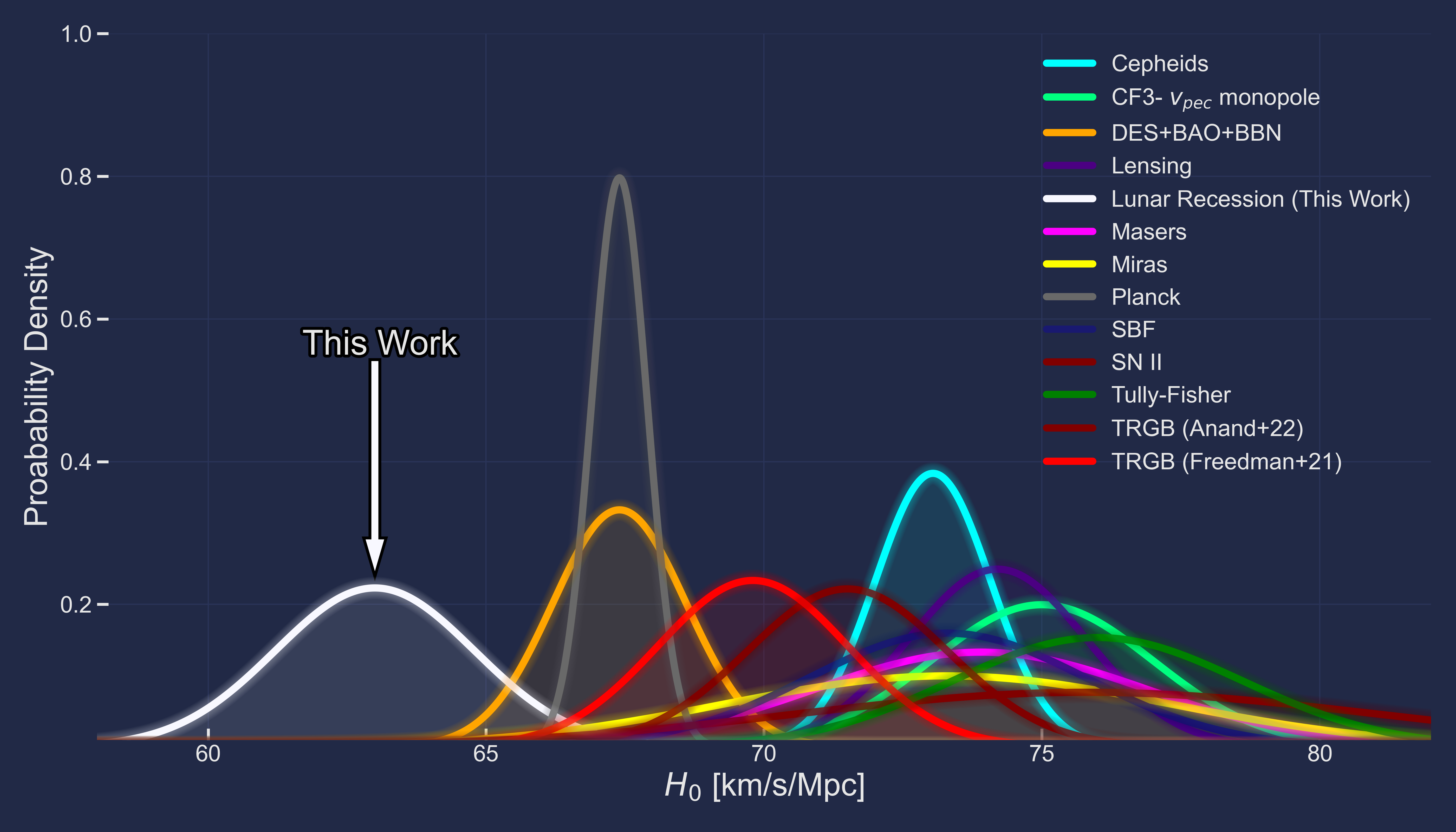}
    \caption{A comparison of numerous recent $H_{0}$ results, including the one we derive in this paper, which we stress is the \textit{only} measurement which is both model-independent and performed in a single step. Methods plotted include Cepheid variables \citep{Riess2022}, a restriction on the peculiar velocity monopole from Cosmicflows-3 \citep{Tully2016}, the Dark Energy Survey + Baryon Acoustic Oscillations + Big Bang Nucelosynthesis \citep{DESBAOBBN}, time delay cosmography of lensed quasars from TDCOSMO \citep{TDCOSMO}, megamasers from the Megamaser Cosmology Project \citep{masercosmo}, Mira variables \citep{Huang2020}, Planck measurements of the CMB \citep{Planck2020}, surface brightness fluctuations \citep{Blakeslee2021}, type II supernovae \citep{dejaeger2020}, the Tully-Fisher relation \citep{Kourkchi2020}, and the Tip of the Red Giant Branch \citep{Freedman2021,Anand2021}.}
    \label{fig:hubble}
\end{figure*}

\section{Method and Data}
We begin with the Hubble law and solve for the Hubble constant:
\begin{equation} \nonumber
    H_0 = \frac{v}{d},
\end{equation}
where $v$ is the recession velocity, and $d$ is the distance from Earth. Scaling by some typical values, we obtain the scaling relation
\begin{equation} \nonumber
    H_0 = 70~\mathrm{km/s/Mpc} \left( \frac{v}{2.755~\mathrm{cm/yr}} \right) \left( \frac{d}{385,000~\mathrm{km}} \right)^{-1}.
\end{equation}
The question, then, is which are the appropriate values of $v$ and $d$ for an estimate of the Hubble constant?

\subsection{Lunar Distance}
Because of the Moon's elliptical orbit, the Earth-Moon distance varies from 363,300 km at perigee to 405,500 km at apogee \citep{moonfacts}. Table~\ref{tab:distances} lists important distance values from the Moon's orbit. Calculating a true cosmic expansion rate from the Moon's recession would require a series of precise, instantaneous measurements of both the Earth-Moon distance and the lunar recession velocity, spanning many orbits. In the absence of that, we will use our best guess. Since we are looking at the Moon's recession over time, we adopt the time-averaged distance of 385,000 km $\pm$ 1 mm as the nominal Earth-Moon distance.

\begin{table}
    \centering
    \vspace{0.5cm}
    \begin{tabular}{l|c}
        \hline
        Distance & Value [km] \\ \hline
        Semimajor Axis & 384,400 \\
        Perigee & 363,300 \\
        Apogee & 405,500 \\
        \textbf{Time-Averaged} & \textbf{385,000} \\ \hline
    \end{tabular}
    \caption{Important distances in the Moon's orbit from \citet{moonfacts}. These distances are precisely determined by laser measurements and have 1 mm precision. Since we consider the Moon's recession over time, we adopt the time-averaged Earth-Moon distance as the appropriate distance.}
    \label{tab:distances}
\end{table}

\subsection{Lunar Recession Velocity}
The value of the lunar recession velocity is the subject of some debate \citep{Poliakow2005, Riofrio2012, Maeder2021}. Laser measurements suggest an instantaneous, present-day recession velocity of 3.8 cm/yr, but this has been regarded by some as anomalously high \citep{Riofrio2012}. Sediment data and eclipse data suggest slower recession speeds of $2.9\pm0.6$ and $2.82\pm0.08$ cm/yr, respectively, and these are supported by detailed numerical simulations \citep{Poliakow2005}. We favor the sediment and eclipse values due to their robustness and agreement with simulations, and we adopt the weighted average of the two values, $v = 2.82 \pm 0.08$ cm/yr.

\subsection{Correction for Tides}
Tides have long been known to influence the Earth-Moon system \citep{1878RSPS...27..419D}, and any measure of the lunar recession must account for this effect. To do this we assume the two effects are entirely independent and each contribute to the total recession linearly. Thus, the cosmological lunar recession $v_\mathrm{cosmo}$ is given simply by: $v_\mathrm{total} = v_\mathrm{tides} + v_\mathrm{cosmo}$. $v_\mathrm{tides}$ varies over time due to a variety of effects and \citet{Poliakow2005} provide a table containing historical values dating back 600 Myr. Taking the $v_\mathrm{tides}$ for the approximate age of the sedimentary data gives $v_\mathrm{tides}=0.34$ cm/yr. This then yields a $v_\mathrm{cosmo}$ value of $2.48$ cm/yr.

\section{Results and Discussion}

Evaluating the Hubble law at the distance and recession velocity of the Moon (after taking into account the effect of tides), we obtain $H_0 =$~\hnot, a 2.8\% determination of the Hubble constant. Figure \ref{fig:hubble} shows our measurement along with several others from the recent literature. We see that previous results which are from the ``late" Universe (e.g. Cepheids) are consistently higher than those from the ``early" Universe (e.g. Planck). Our measurement is important not only because it is the only measurement of the Hubble Constant which is both model-independent and performed in a single step, but because it is the first local measurement of $H_{0}$ whose value is *below* the value measured by \textit{Planck}. 

\section{Summary and Future Outlook}

In this letter, we examine the dynamics of the Earth-Moon system to provide the first truly local measurement of the Hubble constant of $H_{0}$ = \hnot. This is somewhat \textit{lower} than the value measured by the \textit{Planck} satellite, though it is roughly consistent within the errors (see Figure \ref{fig:hubble}). We emphasize that our measurement of $H_{0}$ is the first which is both model-independent and is done within a single step (i.e. does not rely on a multi-step distance ladder). Our work is an important result, as it suggests that other local Universe results whose measurements are higher than those of \textit{Planck} may be biased by systematic errors. 

In the near future, we hope to improve the precision on our lunar-based measurement of $H_{0}$ by both 1) collecting more geological samples (to improve the sedimentary data), and 2) observing in detail more eclipses (to improve the eclipse-based measurements). Both of these avenues will allow us to decrease the uncertainty on our value of $H_{0}$ and solve once and for all this cosmological crisis.

\acknowledgments
The idea for this paper came in 2019 out of late night desperation of three grad students trying to procrastinate real work. Nearly 3 PhDs and some unknown fraction of a pandemic later, we are enthralled to report this very significant result, which is definitely not a joke.

\bibliography{paper}
\bibliographystyle{aasjournal}


\end{document}